\documentclass[prd,tightenlines,nofootinbib,superscriptaddress,twocolumn]{revtex4}

\usepackage[colorlinks=true,linkcolor=black]{hyperref}
\usepackage{graphicx}
\usepackage[utf8]{inputenc}
\usepackage[T2A,T1]{fontenc}	
\usepackage{ae}
\usepackage{aecompl}
\usepackage{textcomp}
\usepackage{amsmath}
\usepackage{amssymb}
\usepackage{amsthm}
\usepackage{amsfonts}
\usepackage{booktabs}
\usepackage{braket}
\usepackage{bbm}
\usepackage[capitalise]{cleveref}
\usepackage{color}
\usepackage{dsfont}
\usepackage[dvipsnames]{xcolor}
\usepackage{esint}
\usepackage{mathrsfs}
\usepackage{multirow}	
\usepackage{refcount}
\usepackage{slashed}
\usepackage{siunitx}
\usepackage{tabularx}
\usepackage{pifont}
\usepackage{xcolor}

\usepackage{todonotes}
\usepackage{tikz}
\usetikzlibrary{external}
\tikzset{external/up to date check=simple}
\usepackage{circuitikz}
\usetikzlibrary{shapes.geometric,calc,intersections,
decorations.markings,decorations.text,decorations,patterns, arrows,
decorations.pathmorphing,fpu,math,babel}

\newcommand{\me}{\mathrm{e}}
\newcommand{\mi}{\mathrm{i}}
\newcommand{\md}{\mathrm{d}}

\DeclareMathOperator{\trace}{tr}

\begin{document}

\title{Einselection from incompatible decoherence channels}

\author{{\bf Alexandre Feller}}\email{alexandre.feller@esa.int}
\author{{\bf Guillaume Coeuret Cauquil}}
\author{{\bf Benjamin Roussel}}\email{benjamin.roussel@esa.int}
\affiliation{%
	Advanced Concepts Team, European Space Agency,
	Noordwijk, 2201 AZ, Netherlands%
}

\begin{abstract}
	Decoherence of quantum systems from entanglement with an unmonitored
	environment is to date the most compelling explanation of the
	emergence of a classical picture from a quantum world. While it is
	well understood for a single Lindblad operator, the role in the
	einselection process of a complex system-environment interaction
	remains to be clarified.  In this paper, we analyze an open quantum
	dynamics inspired by CQED experiments with two non-commuting
	Lindblad operators modeling decoherence in the number basis and
	dissipative decoherence in the coherent state basis.
	We study and solve exactly the problem using
	quantum trajectories and phase-space techniques. The einselection
	optimization problem, which we consider to be about finding 
	states minimizing the variation of some entanglement witness at
	a given energy, is studied numerically. We show that Fock states 
	remain the most robust states to decoherence up to a critical coupling.
\end{abstract}

\maketitle

\section{Introduction}

The strangeness of the quantum world comes from the principle of
superposition and entanglement. Explaining the absence of those
characteristic features in the classical world is the key point of the
quantum-to-classical transition problem.
The fairly recent theoretical and experimental progresses
have largely reshaped our understanding of the emergence
of a classical world from quantum theory alone. 
Indeed, decoherence theory \cite{Zurek_2003}
has greatly clarified how quantum coherence is apparently lost for 
an observer through the entanglement of the system with a 
large unmonitored environment. At the same time, 
this interaction with the environment allows to understand
the emergence of specific classical pointer states, a process
called einselection. 

However, in the case of complex environments, the physics of decoherence
can be much more involved. It is then necessary to consider the structure
of the environment.
A recent approach exploring this path, called quantum Darwinism
\cite{Zurek_2008,Zurek_2009}, considers an environment composed of
elementary fragments accessing a partial information about the system.
While a lot of insights were originally
model-based~\cite{Zwolak-2009,Riedel-2010,Riedel-2011,
Riedel-2012,Riedel-2016}, it has been proved that some basic assumptions
of the quantum Darwinism approach, like the existence of a common
objective observable, come directly from the theoretical framework of
quantum theory~\cite{Brandao-2015,Adesso-2018,Qi_2020}.

Another possible source of complexity comes from
the possibility that the environment can measure different
observables of the system. Formally, the effective dynamics
of the system will be described by many Lindblad operators
which may be non commuting. When right eigenvectors of
a Lindblad operator exist, which correspond to exact pointer
states, we say that the operator describes a decoherence
channel. Non-commuting Lindblad operators will give
rise to what we call incompatible decoherence
channels~\cite{arthurs1965bstj,Briegel1993,torres2014closed,Torres2019}.
Interesting physical effects are coming from the subtle interplay 
between those incompatible channels. For instance, the physics
of a quantum magnetic impurity in a magnetic medium can be
modeled as an open quantum system problem in which the
environment probes all three Pauli matrices, instead of only
one as in the standard spin-boson model~\cite{Neto-2003,Novais-2005}.
Depending on the relative values of the coupling constants,
different classical regimes emerge at low energy with one channel
dominating the others. What's more, for some fine tuned cases,
the more interesting phenomena of decoherence frustration
occurs~\cite{Novais-2005} where all channels contribute
to suppress the loss of coherence of the system at all energy
scales. This kind of subtleties in the einselection process
have also been studied in circuit
QED~\cite{Degiovanni-2008,jordan2005continuous,ruskov2010qubit,hacohen2016quantum,essig2020multiplexed}
where, once again, the emergent classical picture is strongly
dependent on the structure of the environment and the relative
strength of its decoherence channels.

Thus, in the case of several incompatible decoherence channels, the
problem of the emergence of a privileged basis is far from trivial.
In this paper, we analyze the einselection process of a 
system in contact with two competing decoherence channels
inspired by cavity QED experiments. The system 
a mode of the electromagnetic field confined in a high-quality 
cavity. The two sources of decoherence, one in the
coherent state family and the other in the number basis, comes for the 
imperfections of this cavity and the atoms used to probe the 
field that are sent through the cavity. We base our analysis on the 
Lindblad equation to describe the effective open quantum dynamics.
The problem is solved exactly using the quantum trajectory, 
the characteristic function and the quantum channel approaches
allowing to properly characterize the different parameter regimes.
These methods allow to go beyond the formal solution~\cite{torres2014closed,Torres2019}
by offering a clearer physical representation of the
dynamics and the relevant timescales of the the problem.
Still, the einselection problem, which
is about finding the most robust (approximate) states to the
interaction with the environment, has to be solved numerically.
We choose those states as the pure states that minimize the
short-time variation of some entanglement witness (the
purity) with the environment at a given fixed energy.
Depending on the relative values of the coupling constant, 
we find the intuitive result that the einselected states interpolate
between Fock and coherent states. However, we uncover the
remarkable fact that Fock states remain exactly the optimal
einselected states up to a critical value of the couplings that
can be obtained analytically. Finally, in the long-term dynamics, the
einselection process appears to be much more complicated.

The paper is structured as follows. In \cref{sec:reduction}, 
we present a summary of techniques to derive the reduced 
dynamics from the full system-environment unitary evolution. This section
can be skipped if the reader accepts \cref{eq:Lindblad:twochannels}.
The core  results are presented in \cref{sec:setup} where
the model is solved exactly. The einselection process
is analyzed in \Cref{sec:einselection} and we present 
numerical evaluation of the Wigner function of the system which
are used to properly analyze the different einselection regimes. 
We conclude in \cref{sec:disc} by discussing how our analysis 
could be generalized and understood from a more abstract perspective
through the algebraic structure of the jump operators.

\section{Motivations from CQED}
\label{sec:reduction}

By managing to entangle a single mode of the electromagnetic
field and an atom, cavity quantum electrodynamics (CQED) is
a nice experimental setup to explore the foundations of 
quantum theory, quantum information and the physics of the 
quantum to classical transition.

One implementation of CQED~\cite{Haroche_book} uses an electromagnetic
mode trapped in high quality factor mirrors as a system which is probed
by a train of atoms acting as two-level
systems (qubits) conveniently prepared. The setup can work in different regimes
where the qubit is in resonance or not with the field. The off-resonance
functioning mode, also called the dispersive regime, is particularly interesting
since the atoms can be thought as small transparent dielectric medium
with respect to the field with an index of refraction depending on its
state. The atom is able to register some phase information about
the field making it a small measurement device. In fact, a train of atoms
can be thought as a non-destructive measurement device probing the 
number of photons in the field.  Focusing our attention on the field itself,
the atoms have to be considered as part of the environment of the
field (even if it is well controlled by the experimentalist) giving rise
to a decoherence channel having the Fock states as pointer
states~\cite{Haroche_book,Haroche-1999,Haroche-2007}. 
Similar experiments can also now be 
performed on circuit QED platforms~\cite{essig2020multiplexed}.

However, a second source of decoherence exists, not controlled by the 
experimentalist this time, and has its origin in the imperfections of the
cavity. Indeed, while the quality factor is high enough to see
subtle quantum phenomenon, photons can still get lost over time in the
remaining electromagnetic environment. This second decoherence channel
leads to a decoherence of the field mode on the coherent state basis of
the electromagnetic field~\cite{Caldeira-1985}, and introduces dissipation.

\begin{figure}
	\includegraphics{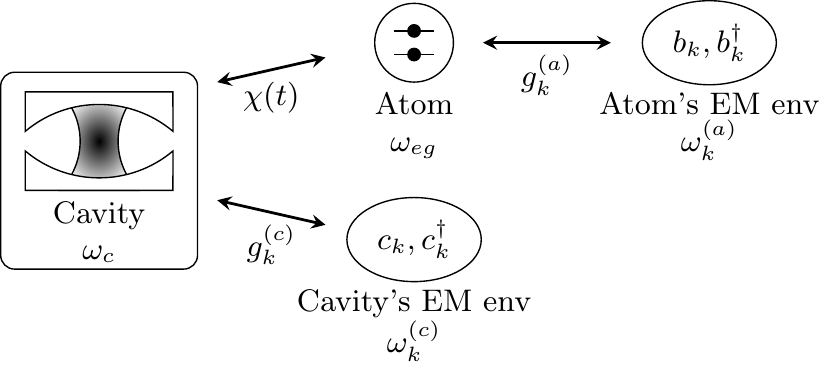}
	\caption{Dynamics of the cavity.
		The cavity is coupled to its own electromagnetic environment and
		to an atom, itself coupled to its own electromagnetic
		environment. We suppose that the two electromagnetic baths are
		not coupled. Eventually, the coupling of the cavity to its bath
		leads to photon losses. The coupling with the atom, in the
		dispersive regime, leads to decoherence on the photon basis.
	}
	\label{fig:cqedmodel}
\end{figure}
We thus have two natural decoherence channels in this experimental
setup, one which selects photon number states and the other coherent
states. However, those two basis are incompatible in the sense that they
are associated to complementary operators, the phase and number operators.
This strong incompatibility between those classical states motivates the 
question on which classical pictures, if any, emerge from such a 
constrained dynamics. 

Before analyzing this problem in details, let's first put the previous
discussion on firmer ground by deriving an open quantum system 
dynamics of the Lindblad form. We use a Born-Markov approximation
of an effective Hamiltonian describing a CQED experiment represented
on \cref{fig:cqedmodel}. As a starting point, consider the following 
Hamiltonian:
\begin{subequations}
\begin{align}
H &= H_S + H_E + H_{SE} \,, \\
H_S &= \hbar \omega_c \, N \,, \\
H_E &= \frac{1}{2} \hbar \omega_{eg} \, \sigma_z + \sum_{k} \hbar \omega^{(c)}_k c^\dagger_k c_k
		+ \sum_{k} \hbar \omega^{(a)}_k b^\dagger_k b_k \nonumber \\
	&+ \sigma_z \cdot \sum_{k} \hbar g^{(a)}_k (b^\dagger_k + b_k) \,, \\
H_{SE} &= \hbar \chi(t) \, N \cdot \sigma_z + a \cdot \sum_k g^{(c)}_k c^\dagger_k + h.c.
\,,
\end{align}
\end{subequations}
with $(a,b,c)$ ladder operators of harmonic oscillators
associated to the system and two environments respectively, 
$N=a^\dagger a$ the number operator for the system oscillator 
$a$ and $\sigma_z$ the $z$ Pauli matrix. 
The system of interest is the mode confined in the cavity. Its free
Hamiltonian $H_S$ is just the free harmonic oscillator at frequency 
$\omega_c$. The cavity mode is coupled to two other systems:
\begin{enumerate}
	\item The atom at frequency $\omega_{eg}$. It is coupled to the
		cavity in the dispersive regime defined as a non zero 
		dephasing $\Delta = \omega_{eg} - \omega_c$
		between the qubit and the cavity. This is the first term 
		of $H_{SE}$. The coupling constant $\chi(t)$ is given
		by $g^2(t)/4\Delta$ where $g(t)$ is the coupling constant in the
		Jaynes-Cummings model. This atom is
		subjected to an electromagnetic environment described by the
		harmonic modes $(\omega^{(a)}_k, g^{(a)}_k)$ coupled to it by a
		spin-boson model. Both the atom and its electromagnetic
		environment form the first decoherence channel.
	\item Other harmonic modes $\omega^{(c)}_k$ describing the
		electromagnetic environment of the cavity. In the secular
		approximation, this environment leads to photon losses in the
		cavity. This is the second decoherence channel.
\end{enumerate}

The first decoherence channel is given by the atom passing through the 
cavity. Its effect on the field is characterized by a correlation function
of the form:
\begin{align}
G^{(a)}_\rho(t,t') = \langle \chi(t)\sigma_z(t)\, \chi(t')\sigma_z(t') \rangle_\rho 
			= \chi(t) \chi(t')
\,.
\end{align}
We remark that, in this case, the correlation function is independent of the state
of the qubit $\rho$. We will rewrite the correlation function in terms of the variable
$\tau = t-t'$.
Supposing that $G^{(a)}_\rho(t,t') $ satisfies the usual assumptions of the Born-Markov
approximation scheme (stationnarity and fast decay), the operators appearing 
in the Born-Markov equation are of the form
$\int_0^{+\infty} G^{(a)}(t,\tau) \, \me^{-\mi \omega \tau} \,\md \tau \, N$.
From there follows a Lindblad equation with the jump operator 
$L_n(\omega) = \sqrt{2G^{(a)}(-\omega)} N$. Unfortunately, it is not possible
to satisfy those requirement for the function $\chi(t)$ (indeed, stationarity implies
that $\chi$ should be a phase but, being real, it must be a constant which cannot satisfy
the decay requirement). Thus, the time dependence must be kept in full generality
in this model, breaking the usual stationarity assumption. This is not a problem
for the derivation of a master equation of the Lindblad form, the only difference being that
the rates will be time dependent. Still making the assumption of fast decay in the
time coordinate $\tau$, the usual steps of the derivation can be followed.
We then end up with a time-dependent Lindblad term proportional to the 
operator $N$: $L_n(t,\omega) = \sqrt{2G^{(a)}(t,-\omega)} N$.

A sufficient condition for the fast decay assumption to be valid is to have an 
exponential decay. This is the case in the Rydberg atoms experiment where
the coupling constant follows the Gaussian beam in the cavity.
If we send a stream of atoms, we still have to make sure that the assumption
of fast decay in $\tau$ is valid. This implies that the atoms must be sent as a
group to have the exponential decay of the correlation function.

The second decoherence channel comes from the leaky cavity and does not 
present any analytical difficulties. Supposing that those degrees of freedom are at
equilibrium and at zero temperature, the Born-Markov equation then reduces to a Lindblad term 
with the jump operator $L_a(\omega) = \sqrt{2G^{(c)}(-\omega)} \, a$
with $G^{(c)}$ the correlation function of the environment field modes
at zero temperature.

Before going to the main analysis of the model, two remarks have to
be made. The first remark is that if we were to prepare 
the experimental setup in the resonant regime $\Delta = 0$, 
the atom-field interaction is modified in such a way that the atom can emit 
or absorb a photon from the mode. Staying in the Markovian regime would 
result in a dissipative dynamics equivalent to a photon emission and absorption 
albeit with very different transition rates. Thus, running the experiment in 
the resonant regime will not result in the dynamics we are interested in.
The second one refers to the system-environment cut and the Markovian
hypothesis. Here, we chose to consider as the system only the field mode,
all the other degrees of freedom then being part of the Markovian 
environment. However, as it was done in~\cite{Degiovanni-2008}, it is  
also possible to consider as the system the field and the qubit while 
all the other electromagnetic  field modes form the Markovian
environment. Different decoherence channels acting on the qubit
and/or on the field can be considered while the non-trivial
internal Jaynes-Cummings dynamics adds another level on
complexity.

\section{Dissipative dynamics}
\label{sec:setup}

\subsection{Position of the problem}
\label{sec:problem}

The previous discussion shows that it is possible in principle
to engineer a complex open quantum dynamics with two 
incompatible decoherence channels. Building on this motivating 
example, the problem we are interested in is to understand the
einselection process of a field mode modeled by a simple
harmonic oscillator subject to two decoherence channels: one
is induced by the dispersive interaction with a train of atoms 
leading to decoherence on Fock states and the other is 
induced by loss of photons in the cavity leading to decoherence
on coherent states.
Abstracting ourselves from the CQED context,  we consider from
now on an open quantum dynamics modeled by a Lindblad equation
with the Hamiltonian $H_S$ and two time-independent jump operators:
\begin{align}
L_a = \sqrt{\kappa_a} \, a 
\quad 
L_n = \sqrt{\kappa_n} \, a^\dagger a =\sqrt{\kappa_n} \, N 
\,.
\end{align}
Those two operators do not commute with each others and therefore
can't be simultaneously diagonalised. The incompatibility between
decoherence channels we are referring to has to be understood in this sense.
Still, their commutator $[L_n,L_a] \propto L_a$ remains simple enough 
and this is the key to find an exact solution~\cite{torres2014closed}. 

Given the two quantum jumps $L_a$ and $L_n$, the Lindblad equation
we want to analyze is:
\begin{align}
\partial_t \rho = 
	- \mi \omega_c\, [N,\rho] 
	&+ \kappa_n \left(N\rho N - \frac{1}{2}\{N^2,\rho\}\right) \nonumber \\
	&+ \kappa_a \left(a\rho a^\dagger - \frac{1}{2}\{N,\rho\}\right)
\,,
\label{eq:Lindblad:twochannels}
\end{align}
where $N$ is the number operator and $\omega_c$ the frequency
of the field mode.

To gain some intuition on the physics behind this dynamics, let's study
the evolution of some average values. We will focus on the average
position in phase space, which can be recovered directly from the 
average value of the annihilation operator $\braket{a}$, as well 
as the variance around this position which are related to the
average photon number $\braket{N}$ and the square of the annihilation
operator $\braket{a^2}$. From those values, we can extract the average
position and the fluctuations on an arbitrary axis 
$\hat x_\theta = \cos\theta \, \hat x + \sin\theta \,  \hat p$.

The observable $N$ is in this case of particular interest. First of all,
it gives access to the average energy of the cavity. Second, it is
not affected by the proper dynamics of the cavity nor by the $L_n$
jumps. A direct computation gives the standard exponential decay of 
a damped harmonic oscillator with rate $\kappa_a$:
\begin{equation}
\langle N(t) \rangle  = \me^{-\kappa_a t} \langle N(0) \rangle
\,.
\end{equation}
This solutions gives the intuitive result that photon loss induces a
decrease of the average number of photons in the cavity. 

On the contrary, both decoherence channels affect the average
values of the annihilation operator $a$ and its square:
\begin{align}
	\langle a(t)   \rangle  
	& = \me^{-\mi \omega_c t} \me^{- \kappa_n t/2} \me^{- \kappa_a t/2} \,
	\langle a(0) \rangle \,, \\
	\langle a^2(t) \rangle
	& = \me^{-2\mi \omega_ct} \me^{-2 \kappa_n t} \me^{- \kappa_a t} \,
	\langle a^2(0) \rangle \,.
\end{align}
We see that the average position in phase space oscillates at the
frequency of the cavity and is exponentially suppressed at a
rate $(\kappa_a + \kappa_n)/2$, different than the decay rate of the
average energy. We expect that, for some states, this separation of
timescales will lead to an increase of fluctuations at intermediate
times.

To put this statement on firmer ground, it is instructive to compute the 
average fluctuations along an axis $x_\theta$ making an angle 
$\theta$ with the horizontal in phase space. A direct computation 
gives:
\begin{align}
	\Delta x_\theta^2(t) = \frac{1}{2}
	&+ \left(\langle N(t) \rangle - |\langle a(t) \rangle|^2 \right) \nonumber \\
	&+ \text{Re} \left[
		\left(\langle a^2(t) \rangle - \langle a(t) \rangle^2\right)
		\me^{-2\mi \theta}
	\right]
\,.
\end{align}
The first term $1/2$ remains even when the cavity is empty and corresponds to
the fluctuations of the vacuum. It is also the minimal isotropic fluctuations 
that satisfy the Heisenberg principle. The second term corresponds 
to isotropic fluctuations above the vacuum state and are, for instance,
zero for coherent states but not for a thermal density matrix. Finally, the last
term is the anisotropic part of the fluctuations and can be either positive or 
negative. Even though the overall fluctuations must satisfy
Heisenberg inequalities, states can exhibit fluctuations below the vacuum
ones in some directions. This property is called squeezing and is of
particular interest for quantum technologies.

If the initial state at $t=0$ is a Fock state $\ket{n_0}$, only the
isotropic contribution $\braket{N(t)} = n_0 \,\me^{-\kappa_a t}$ survives.
In this case, the fluctuations are decreased solely because of the loss
of energy in the cavity. This is due to the fact that the $L_a$ channel
does not create any coherences between different Fock states, while the
$L_n$ channel does not affect statistical mixtures of Fock states.

However, the situation is more subtle if we prepare a coherent state.
A coherent state $\ket{\alpha}$ such that $\alpha
= \sqrt{n_0} \me^{\mi \varphi}$ gives the values $\braket{N(0)} = n_0$ and
$\braket{a^2(0)} = n_0 \me^{2 \mi \varphi}$. The fluctuations for such a
state are evolving according to:
\begin{multline}
\Delta x_\theta^2(t) 
	= \frac{1}{2} + \\
	n_0 \me^{-\kappa_a t} (1 - \me^{-\kappa_n t}) \left[
		1 - \me^{-\kappa_n t}
		\cos\left(2(\varphi - \theta - \omega_c t)\right)
	\right]
\,.
\label{eq:fluc_coherent}
\end{multline}
As in the Fock state case, the fluctuations are exponentially suppressed
in time as the cavity looses energy with a rate $\kappa_a$. However, the
situation is more interesting, because of the competition between both
channels on the isotropic part of the fluctuations: on top of the
exponential suppression, the $L_n$ channel tends to increase
fluctuations up to the maximum possible for the average energy still
stored in the cavity. As we will see later, this comes from the fact
that the coherences between different photon numbers are suppressed over
a timescale $1/\kappa_n$.

The anisotropic part of the fluctuations, in this case, is always
smaller than the isotropic part over the vacuum. The fluctuations will
thus stay always above the vacuum fluctuations in any directions: this
dynamics does not exhibit squeezing. At short times, however, the
fluctuations behave as $\Delta x^2_\theta(t) \simeq 1/2 + 2 n_0 \kappa_n t
\sin^2(\varphi - \theta - \omega_c t)$. They stay identical in the
axis given by $(0,\alpha)$, but are increased with a rate $n_0 \kappa_n$
in the orthogonal direction.  At longer times, the anisotropic
fluctuations decay, with an exponential rate~$(\kappa_a + \kappa_n)$,
faster than the isotropic decay rate~$\kappa_a$.

Those timescales are very important to properly follow the einselection
process and we will analyze them in detail after having obtained the general
solution.

\subsection{Quantum trajectory approach}
\label{sec:trajectory}

The Lindblad equation~\labelcref{eq:Lindblad:twochannels}
can be solved exactly from the quantum trajectory approach~\cite{Dalibar_1992}.
One of the motivations for using this approach is that it is now
possible to explore the dynamics of well-controlled quantum systems at
the level of a single experimental realization~\cite{Ficheux_2018}. 
In this sense, it is closer to the latest experiments studying 
decoherence. 
The strategy will be to first write the stochastic Schr\"odinger
equation for the system, parametrise properly a trajectory,
write its relative state and finally average over all possible 
trajectories to obtain the reduced density matrix.

The stochastic Schrödinger equation associated to the 
Lindblad \cref{eq:Lindblad:twochannels} can be written as:
\begin{multline}
\ket{\psi_c(t+\md t,[\mu])} =\\
\begin{cases}
	\left(
		\mathbbm{1}
		- \md t \left[
			\mi \omega  a^\dagger a
			+ \frac{\kappa_a}{2} a^\dagger 
			a+\frac{\kappa_n}{2} (a^\dagger a)^2
		\right]
	\right)
	\ket{\psi_c(t,[\mu])}
	\\
	\hfill\text{if $\mu(t)=0$,}\\
	-\mi \sqrt{\md t \kappa_n}\, a^\dagger a
	\ket{\psi_c(t,[\mu])}
	\hfill\text{if $\mu(t)=1$,}\\
	-\mi \sqrt{\md t \kappa_a}\, a
	\ket{\psi_c(t,[\mu])}
	\hfill\text{if $\mu(t)=2$.}
\end{cases}
\label{eq:stochaschro}
\end{multline}
where $\mu(t)$ takes the values $0$, $1$, and $2$ if there is no quantum
jump, a jump $L_a$ or a jump $L_n$ respectively. A trajectory will
be then parametrized by the type of jump and their occurrence time. We
stress that the state $\ket{\psi_c(t, [\mu])}$ is not normalized, the
probability of the trajectory $t \mapsto \mu(t)$ being given by $p[\mu] =
\braket{\psi_c(t,[\mu]) | \psi_c(t,[\mu])}$.

\begin{figure}[h!]
	\centering
	\includegraphics{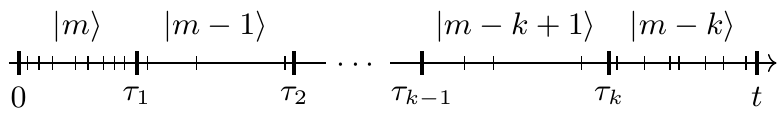}
	
	\caption{Parametrization of a quantum trajectory build from 
	two types of quantum jumps $L_a = \sqrt{\kappa_a}\, a$ and 
	$L_n = \sqrt{\kappa_n} \, a^\dagger a$. The times $\tau_\sigma$ 
	correspond to the $L_a$ jumps and the quantum state 
	between each jump is above each slice.
	}%
	\label{fig:division}
\end{figure}

To have a better intuition of how a state evolves, let's start from a Fock
state $\ket{m}$. Since it is an eigenstate of $a^\dagger a$, the
trajectories $\mu=0$ and $\mu=1$ do not change the state and only induce a
phase. However, the jump~$\mu=2$ induces a photon loss and the state is
changed into $\ket{m-1}$ with a phase.
Thus, a proper parametrization of a trajectory is to slice the evolution 
according to the $L_a$ jumps, as shown in \cref{fig:division}. There, 
a trajectory is parametrized by:
\begin{itemize}
	\item $N_a$ jumps $L_a$, indexed by the letter $\sigma$, occurring
		at times $\tau_\sigma$,
	\item $N_\sigma$ jumps $L_n$ in the slice $\sigma$, indexed by the
		letter $s$, occurring at times $t_{\sigma,s} \in ] \tau_\sigma,
		\tau_{\sigma + 1}[$.
\end{itemize}

For a given slice $\sigma \in [\![1,k+1]\!]$ where $\tau_{k+1}=t$ and
$\tau_0=0$, the relative state obtained from \cref{eq:stochaschro} is
proportional to the Fock state $\ket{m-\sigma}$. By denoting
$\alpha(t,[\mu])$ the proportionality constant, we find:
\begin{align}
&\ket{\psi_c(\tau_\sigma,[\mu])}
	=\prod_{s=1}^{N_{\sigma-1}}
	\left(-\mi \sqrt{\kappa_n \md t_{\sigma-1,s}}(m-\sigma+1)\right) \nonumber \\
	&\times \left(
		-\mi \sqrt{\kappa_a \md \tau_{\sigma}(m-\sigma+1)}
	\right) \nonumber \\
	&
	\times \me^{%
		-\left(
			\mi (m-\sigma+1) \omega_c
			+ (m-\sigma+1)^2 \kappa_n/2
			+ (m-\sigma+1) \kappa_a/2
		\right) 
		(\tau_{\sigma}-\tau_{\sigma-1})
	} \nonumber \\
	&\times \alpha(\tau_{\sigma-1}, [\mu]) \ket{m-\sigma}
\,.
\label{eq:fock_state_2deco}
\end{align}

Having the formal form of the state of the system relative to 
a given quantum trajectory, we can obtain the reduced density 
matrix by summing over all possible trajectories. Our parametrization
is such that we can re-sum all the phases accumulated between each
$L_a$ jumps easily and then average over all $L_a$ jump events.
For the initial Fock state $\ket{m}$, only the diagonal elements of the 
reduced density matrix can be non zero. No coherences are induced 
since a Fock state remains a Fock state and no superposition appears.
This immensely simplifying feature comes directly from the special
commutation relation of the jump operators. We obtain:
\begin{align}
	\rho_{m-k,m-k}
	=
	{m\choose k} (1- \me^{-\kappa_a t})^k \me^{-\kappa_a t(m-k)}
	\,.
	\label{eq:fock:proba}
\end{align}
Starting from a Fock state $\ket{m}$, the state $\ket{m-k}$ is reached 
if $k$ photons have leaked. The associated probability is given by
$
 p(t,k) = {m\choose k} (1- \me^{-\kappa_a t})^k \me^{-\kappa_at(m-k)}
 $,  
which is the classical probability for a binomial experience $B(m,p)$,
where the probability $p$ for a photon to leak between $0$ and $t$ is $p
=1- \me^{-\kappa_a t}$. We recover the standard evolution of a
damped harmonic oscillator prepared in a Fock state.

The coupling constant $\kappa_n$ does not appear in this expression
which means that the dephasing induced by an atom flux has no effect
on a Fock state. Indeed, we saw that $L_n$ jumps
induce a dephasing between Fock states. Since a unique trajectory keeps
the cavity in a single Fock state, this dephasing is only a global
phase shift and thus has no effect at all. 

Since Fock states form a basis of the Hilbert space of the system, an
analogous computation starting from any superposition of Fock states
gives the general solution:
\begin{align}
	\rho_{m,n}(t)&= 
	\me^{-\mi \omega (m-n) t}
	\me^{-\kappa_a\frac{m+n}{2}t}
	\me^{-\frac{\kappa_n}{2}(m-n)^2 t} \nonumber \\
 	&\times\sum_{N_a} \sqrt{{m+N_a\choose N_a}{n+N_a\choose N_a}} 
 		(1- \me^{-\kappa_a t})^{N_a} \nonumber\\
	&\times \rho_{m+N_a,n+N_a}(0) 
\,.
\label{eq:density_matrix_gen_2deco}
\end{align}

The physical content of this general exact expression is more transparent
if we initially prepare a superposition of coherent states. What's
more, those are the typical quantum states that are used in 
cavity quantum electrodynamics to study the decoherence 
process. Let's then consider the state 
$
\ket{\Psi_c}
=
\frac{\ket{\alpha_+}+\ket{\alpha_-}}{\sqrt{\mathcal{N}}}
$
where 
$\alpha_\pm=\alpha \, \me^{\pm\mi \theta/2}$,
$\alpha\in\mathbb{C}$ and $\mathcal{N}$ is a
normalization factor. The initial density matrix is then given by:
\begin{align}
\rho_{m,n}(0) 
	&= \frac{1}{\mathcal{N}}
	\big(
		\rho_{m,n}(\alpha_+) + \rho_{m,n}(\alpha_-) \nonumber \\
		&+\rho_{m,n}(\alpha_+,\alpha_-) +\rho_{m,n}(\alpha_-,\alpha_+)
	\big)
\,,
\end{align}
where $\rho_{m,n}(\alpha)=\langle m |\alpha\rangle\langle\alpha|
n\rangle$ refers to the matrix element
$mn$ in the Fock basis of the density matrix of the coherent state $\ket{\alpha}$.
The last two terms correspond to interferences between the two coherent
states, with $\rho_{m,n}(\alpha_+,\alpha_-) = \langle m
|\alpha_+\rangle\langle\alpha_-| n\rangle$. Using
\cref{eq:density_matrix_gen_2deco}, the time-evolved density matrix at
time $t$ is given by:
\begin{multline}
\rho_{m,n}(t)
	=\frac{1}{\mathcal{N}}\me^{- \kappa_n \frac{(m - n)^2}{2}t}
	\big[
		\rho_{m,n}(\alpha_+(t))
		+ \rho_{m,n}(\alpha_-(t)) \\
		+ \me^{-|\alpha|^2 (1-\me^{-\kappa_a t})(1-\me^{\mi \theta})}
		\rho_{m,n}(\alpha_+(t),\alpha_-(t))
		+\text{h.c.}
	\big]
\,,
\label{eq:solution:coherentstate}
\end{multline}
where
$\alpha_{\pm}(t) =
\alpha_{\pm} \, \me^{-\mi \omega_c t}  \me^{-\frac{\kappa_a}{2}t}$. 
The effect of the two decoherence channels can be clearly identified. 
First, the term 
$d_a(t) = \me^{-|\alpha|^2 (1-\me^{-\kappa_a t})(1-\me^{\mi \theta})}$
is the usual decoherence factor coming from the jump $L_a = \sqrt{\kappa_a} a$ 
(damped harmonic oscillator) which tends to destroy coherences
between coherent states. On the other hand, the term
$\me^{- \kappa_n \frac{(m - n)^2}{2}t}$ 
tends to destroy coherences between Fock states which are the 
natural pointer states of the channel $L_n = \sqrt{\kappa_n} N$.

The interesting and surprising aspect of \cref{eq:solution:coherentstate}
is that, while the Lindblad jump operators do not commute, the overall
evolution is, in a sense, decoupled. This can be made precise by using the 
quantum channels perspective. Indeed, by considering the superoperators
$\boldsymbol{H} = -\mi ( \bar{H}\otimes \mathbbm{1} - \mathbbm{1}\otimes H)$
and 
$\boldsymbol{D} = \sum_\mu \bar{L}_\mu \otimes L_\mu - \frac{1}{2} \mathbbm{1} \otimes L_\mu^\dagger L_\mu
- \frac{1}{2}  \bar{L}_\mu^\dagger \bar{L}_\mu \otimes \mathbbm{1}$ 
(the bar notation corresponds to complex conjugation), the Lindbald
equation can be formally 
integrated as~\cite{Havel-2003}:
\begin{align}
\rho(t) = \me^{ (\boldsymbol{H} + \boldsymbol{D}) t } \rho(0) 
= \mathcal{L}_{H,[L_\mu]}(t) \rho(0)
\,.
\end{align}
The notation $\mathcal{L}_{H,[L_\mu]}(t)$ is here to remind us that the
quantum channel depends on the set of jump operators $L_\mu$
through the operator $\mathcal{D}$. In our problem, we have two
non-commuting jump operators $L_a = \sqrt{\kappa_a} \, a $ and $L_n =
\sqrt{\kappa_n} N$. Astonishingly,
their respective quantum channels do commute, leading to:
\begin{align}
	\mathcal{L}_{H,L_a,L_N}(t) = \mathcal{L}_{H,L_a}(t) \mathcal{L}_{L_N}(t) 
	= \mathcal{L}_{L_N}(t) \mathcal{L}_{H,L_a}(t)
\,. 
\end{align}
Thus, while the generators of the open quantum dynamics do not commute,
the associated quantum channels do (and decouple in this sense). 
The full dynamics of the model can be solved by parts by looking
at the evolution of both quantum channels separately, which is an easy
task. Nevertheless, this does not trivialize the einselection
problem of finding the exact or approximate pointer states that entangle
the least with the environment. The emergence of a classical picture
does depend on the ``fine structure'' of the dynamics.

\begin{table*}
\caption{Comparison of the timescales relative to the classes of exact pointer states}
	\label{table:timescales}
\centering
	\def\cmark{\ding{51}}
	\def\xmark{\ding{55}}
	\renewcommand*{\arraystretch}{1.1}
	\setlength{\tabcolsep}{8pt}
\begin{tabular}{@{}llcc@{}}
\toprule
	\multicolumn{1}{c}{\textbf{Timescale}} &
	\multicolumn{1}{c}{\textbf{Expression}} &
	\textbf{Coherent state (C)} &
	\textbf{Fock state (F)}\\ 
\midrule
	Decoherence of coherence states &
	$\tau_a = 1/2 \, \langle n\rangle
\kappa_a\sin^2(\theta/2)$ & Decoherence (C) &  Spreading \\
	Relaxation &
	$\tau_r = 1/\kappa_a$ & Relaxation  & Relaxation \& Decoherence (C)\\
	Spreading in angle &
	$\tau_s = 2/\kappa_n \langle n \rangle$ & Spreading & Decoherence (F) \\
	Crown formation &
	$\tau_c = 1/\kappa_n$ & Decoherence (F)& \xmark \\
\bottomrule
\end{tabular}
\end{table*}

Two natural decoherence timescales can be defined, still considering the 
initial state to be the superposition of coherent states $\ket{\Psi_c}$. 
The first one is the usual decoherence timescale $\tau_a$
of a superposition of coherent states induced by the decoherence
channel $L_a$. Its expression is directly given by the
exponential modulation of the coherence term in the general
solution~\cref{eq:solution:coherentstate}:
\begin{align}
\tau_a = \frac{1}{2\,\langle n\rangle \,\kappa_a\,\sin^2(\theta/2)}
\,,
\label{eq:timescales:taua}
\end{align}
using that $ |\alpha|^2 = \langle n\rangle $. Over time, this channel
tends to empty the cavity. The relaxation timescale $\tau_r$ deduced
from the equation $\alpha(t) = \me^{-\kappa_a t/2} \alpha$ is such
that:
\begin{align}
\tau_r = \frac{1}{\kappa_a}
\,.
\end{align} 

A second decoherence timescale $\tau_n$ is associated to the $L_n$ 
channel. In the Fock basis, its expression is straightforwardly given by
$\tau_s = 2 / \kappa_n (m - n)^2$. Still, we can rewrite it in a form 
more suitable for coherent states.
Indeed, coherent states have a 
Poissonian distribution of photons with an average
$\langle n \rangle = |\alpha|^2$ and a variance $\Delta n =
\sqrt{|\alpha|^2} = \sqrt{\langle n \rangle}$.  Thus, the characteristic
width $\Delta n$ of a coherent state gives a characteristic upper bound
$m-n \leq \Delta n$. This allows us to extract a characteristic time
which, as we will see, correspond to a spreading in the angle variable
in phase space:
\begin{align}
\tau_s =\frac{2}{\kappa_n \langle n \rangle}
\,.
\end{align} 
This timescale corresponds to the short time effect of $L_n$. The long
time effect is given for $m-n = 1$ and corresponds to a disappearance 
of the Fock states superposition. The associated timescale $\tau_c$, 
which corresponds to the formation of a rotation-invariant crown in
phase space, is given from \cref{eq:solution:coherentstate} by:
\begin{align}
\tau_{c} = \frac{1}{\kappa_n}
\,.
\end{align}
Thus, the two natural decoherence timescales for coherent
states are $\tau_a$ and $\tau_c$, corresponding to the loss of 
coherence on the coherent state and number state basis respectively.
The scaling $\tau_a / \tau_c \simeq \kappa_n/\kappa_a \langle n\rangle$
implies that for sufficiently high energy, we can have a separation
of decoherence times, by first seeing a decoherence over the coherent
states followed by one on the number basis. \Cref{table:timescales} 
summarizes the different timescales, their expression 
and interpretation when a superposition of coherent states is initially prepared.
However, as we will see over the next sections, the interpretation
of those timescales is relative to the class of pointer states we 
use like, for instance, when we initially prepare
Fock states.

\subsection{Phase-space approach}
\label{sec:characteristic}

\subsubsection{The general solution}

Last section presented the general solution of the Lindblad
\cref{eq:Lindblad:twochannels} using the quantum trajectories
approach. It is also possible to solve the same problem
from a phase-space perspective using characteristic functions
of the density matrix. This approach offers interesting physical 
insights on the dynamical evolution imposed by the two 
incompatible channels.

Given a density operator $\rho$, we associate a function
$C_\rho(\lambda,\lambda^*)$, with $\lambda\in\mathbb{C}$, defined as:
\begin{align}
	C_\rho(\lambda,\lambda^*)
	=
	\trace \left( \rho \me^{\lambda b^\dagger} \me^{-\lambda^* b} \right)
\,.
\label{eq:chara:def}
\end{align}
This function is called the characteristic function adapted to the normal
order. All the normally-ordered average values can be recovered from it.
It is an integral transform of the Wigner function that we will
define and use extensively latter on.
Now, by taking each term of the Lindblad \cref{eq:Lindblad:twochannels}, 
it is a straightforward computation, using the BCH formula, to write it 
in terms of the characteristic function as:
\begin{multline}
	\partial_t C_\rho
	+ \lambda \left(\frac{\kappa_a}{2} - \mi \omega_c\right)
	\, \partial_\lambda C_\rho
	+ \lambda^* \left(\frac{\kappa_a}{2} + \mi \omega_c\right)
	\, \partial_{\lambda^*} C_\rho \\
	=
	-\frac{\kappa_n}{2}
	(\lambda \partial_\lambda - \lambda^*\partial_{\lambda^*})^2
	\, C_\rho
	\,.
	\label{eq:Lindblad:charafunction:twochannels}
\end{multline}
By using polar coordinates $\lambda = r \, \me^{\mi \theta}$, we obtain the 
more transparent form:
\begin{align}
	\left(
		\partial_t
		+ \frac{\kappa_a}{2} r \partial_r
		- \omega_c \partial_\theta
	\right) C_\rho(r,\theta,t)
	= 
	\frac{\kappa_n}{2}
	\partial^2_\theta C_\rho(r,\theta,t) 
	\,.
\end{align}
This equation is of the Fokker-Planck type with the right-hand side
being a diffusion term with a diffusion coefficient given by the
coupling constant $\kappa_n$. Thus, the channel $L_n$ tends 
to spread in angle the characteristic function. This is in accord with the
first discussion of the evolution of fluctuations given through 
the average values. Note that we could even consider an
environment at finite temperature and obtain a similar equation
(with an inhomogeneous term) that can also be solved exactly.

When only the  channel $L_a$ is present, the differential equation 
is first order and can  be solved by the method of
characteristics~\cite{carmichael1999statistical,Arnold-2013}.
This is not directly the case here but can be remedied by going to the
Fourier domain with respect to the variable $\theta$.
Doing this is indeed quite intuitive if we remember that the 
conjugated observable associated to $\theta$ is the number 
operator $N$ which is the natural observable of the problem,
and that both are related by a Fourier transform.

Let's then define a new characteristic function $C_\rho(r, n) = \int C_\rho(r,\theta) \,
\me^{\mi n \theta} \, \md \theta$ which is the Fourier series of 
$C_\rho(r,\theta)$. We then end up with an in-homogeneous 
first-order partial differential equation:
\begin{align}
	\left(
		\partial_t
		+ \frac{\kappa_a}{2} r \partial_r
	\right) C_\rho(r,n,t)
	= 
	\left( \mi \omega_c n - n^2 \right)
	\frac{\kappa_n}{2} C_\rho(r,n,t) 
	\,.
\end{align}
By applying the method of characteristics, solving the equations
$\dot{r} = \kappa_a$ and $\dot{n} = 0$, we obtain the general
solution for an initial condition $C_0(r_0,n_0)$:
\begin{align}
	C_\rho(r,n,t)
	=
	C_0(r\me^{-\kappa_at/2},n)
	\, \me^{\left( \mi \omega_c n  - \frac{\kappa_n}{2} n^2 \right) t }
\,.
\label{eq:chara:solution}
\end{align}
From the radial damping, we recover the usual damping rate leading
to decoherence on the coherent state basis. Besides this usual 
behavior, we have a new damping term depending on the square of
the Fock variable $n$ which leads to the decoherence in the Fock 
basis with the characteristic angle-spreading timescale $\tau_s$ already
uncovered  in \cref{sec:trajectory}.

\subsubsection{Fock state decoherence}

Different initial states can be prepared. We will naturally focus 
on two classes of states, Fock states $\ket{n}$ and coherent
states $\ket{\alpha}$. As a warm up example, suppose that only the
channel $L_n$ is present. If we prepare a Fock state $\ket{n}$, whose
characteristic function is given by 
$C_{\ket{n}}(\lambda) = \mathcal{L}_n( |\lambda|^2)$
where $\mathcal{L}_n$ is the Laguerre polynomial of order
$n$, we know that it will not be affected by the environment and will 
evolve freely as can be readily checked from \cref{eq:chara:solution}.
The important remark here is that the phase-space representation 
of a Fock state or any statistical mixture of them is rotation invariant.

If we now prepare a quantum superposition of Fock states
$(\ket{n_0} + \ket{n_1})/\sqrt{2}$, it is straighforward to 
show that the coherence part of the characteristic function
${C_{01}(\lambda,\lambda^*) = \trace \left( \ket{n_0}\!\!\bra{n_1} 
\me^{\lambda b^\dagger} \me^{-\lambda^* b} \right)}$ evolves as:
\begin{align*}
C_{01}(\lambda,\lambda^*)
	= \me^{- \frac{\kappa_n}{2} (n_1-n_0)^2 t } 
		C_{01}(r,(\omega_c - \theta)t)
\,.
\end{align*}
The coherence term between Fock states are damped by an
exponential factor with a characteristic timescale 
$2/(n_0-n_1)^2 \kappa_n$ scaling as the quadratic inverse
of the ``distance'' between the two components of the
state. The is the expected decoherence dynamics for the
exact pointer states of a decoherence channel.

\subsubsection{Coherent states and the Wigner representation}

In the spirit of cavity quantum electrodynamics experiments, it 
is more appropriate to study the evolution of coherent states
and their superposition. Before discussing the dynamics of such
states, it is necessary to introduce a better-suited phase-space
representation then the one defined by \cref{eq:chara:def}.
Indeed, this latter function is a complex-valued function which is not 
the best choice for representation purposes. From the 
characteristic function $C_\rho(\lambda,\lambda^*)$, we can 
recover an equivalent, real-valued, phase-space representation
called the Wigner function $W_\rho(\alpha)$ of the state $\rho$
(with the parameter $\alpha\in\mathbb{C}$) defined as the 
following integral
transform~\cite{wigner1932quantum,carmichael1999statistical}:
\begin{align}
W_\rho(\alpha) = \frac{1}{\pi^2} \int C_\rho(\lambda) \, \me^{-|\lambda|^2/2}
	\, \me^{\alpha\lambda^*-\alpha^*\lambda}
	\, \md^2 \lambda
\,.
\end{align}
Using the position and momentum coordinates in phase
space $\alpha = x + \mi p$, we recover the common expression:
\begin{align}
W_\rho(x,p) = \int \langle x + u/2 | \rho | x-u/2 \rangle 
	\, \me^{-\mi u p}
	\, \frac{\md u}{2 \pi}
\,,
\end{align}
where the states $\ket{x}$ are eigenstates of the position
operator $x \propto a + a^\dagger$.
The Wigner function is used in a wider context than quantum optics
\cite{ville1948theorie,Flandrin-1998} and possesses a nice set of
properties to represent
quantum interferences in a transparent way. We will use it 
throughout this paper to represent our analytical and numerical
results. Furthermore, since the proper dynamics of the field mode can be
factored out, the dynamics is pictured without the rotation in phase
space that it introduces.

Both for the numerical calculations as well as for understanding the
dynamics in phase space, it is helpful to write the Wigner function in
terms of its Fourier components of the polar angle~$\theta$. Thus, let's
decompose the Wigner function on the Fock state basis
as:
\begin{align}
	W(x,p) &= \sum_{m,n} \rho_{mn} W_{mn} \,, \\
	W_{mn}(x,p) &= \int
		\Braket{x + \chi/2 | m}
		\Braket{n | x - \chi/2}
		\me^{- \mi p \chi}
		\, \frac{\md \chi}{2\pi}
\,.
\end{align}
If we denote $x + \mi p = r \me^{\mi \theta}$, we find that for $l \ge
0$:
\begin{equation}
	W_{k,k+l}(r,\theta)
	=
	\frac{(-1)^k}{\pi}
	\sqrt{2^l \frac{k!}{(k+l)!}}
	\me^{-\mi l \theta}
	r^k
	L_k^l(2 r^2)\,,
\end{equation}
where $L_m^k$ are generalized Laguerre polynomials.  This implies that
the coherence given by $\rho_{k, k+l}$ ($k \in \mathbb{N}, l \ge 0$) and
$\rho_{k-l, k}$ ($k \in \mathbb{N}, l \le 0$) corresponds to the $l$th
harmonics of the angle variable in the Wigner function.

Notably, since the $L_n$ channel acts by multiplying all the elements
situated at a distance~$l$ from the diagonal by the same quantity, it
corresponds to reduce the different harmonics of the angle variable by
an exponential amount. This corresponds of the idea of a spread in the
angle variable.

If we now prepare a coherent state $\ket{\alpha}$ whose characteristic
function is given by
$C_{\ket{\alpha}}(\lambda) = \me^{\alpha^*\lambda-\alpha\lambda^*}$,
its evolution in the presence of only the $L_n$ decoherence channel is given
by the convolution of the initial characteristic function with a Gaussian
function $\mathcal{N}_\sigma(\theta)=
\me^{-\theta^2/2\sigma^2}/\sqrt{2\pi\sigma^2}$ 
of the angle variable $\theta$ spreading in time with a variance
$\sigma^2 = \kappa_n t$ (see \cref{app:Fock}). In terms of the
Wigner function, we have:
\begin{align}
W(r,\theta,t) =
	\int_{-\infty}^{+\infty}
		W_0(r,\theta') \, \mathcal{N}_{\sigma^2 = \kappa_n t}\left(
		\theta - \theta' 
	\right)
	\, \md \theta'
\,.
\label{eq:wigner:fockdeco}
\end{align}
with $W_0$ the initial Wigner function. This form displays clearly the diffusive
dynamics induced by the $L_n$ decoherence channel in the phase angle, putting
the initial intuition on firm ground.

\begin{figure}
	\centering
	\includegraphics{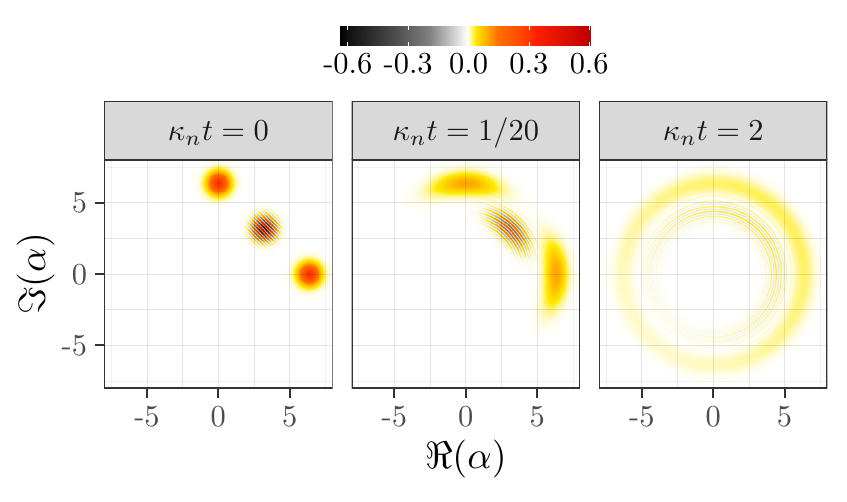}
	\caption{%
		Evolution of a superposition of coherent states from $t=0$, subject to the 
		$L_n$ channel only ($\kappa_a = 0$)
		in the Wigner function representation.
		The Gaussian spots are spread over a timescale given by 
		$\kappa_n \tau_s = 2/ \langle n \rangle$. For the chosen parameters
		$\alpha = \sqrt{40}$, it is given by $\kappa_n \tau_s = 1/20$. At
		longer times $\kappa_n t=2$, the superposition evolves
		toward a statistical mixture of Fock states with the characteristic
		rotation invariance in phase space.
	}
	\label{fig:wigner_fock}
\end{figure}

\Cref{fig:wigner_fock} represents the evolution of the Wigner function 
of a superposition of coherent states.
At a time scale $\tau_s = 2/\kappa_n \langle n \rangle$, the Gaussian 
spots start to spread along circles with a radius given by their respective 
amplitude. The same can be said for the Gaussian interference spot which is 
centered at the mid-point in phase space. At later times $t \gtrsim \tau_c$,
the coherent states spread uniformly and are completely decohered as a
statistical mixture of Fock states\footnote{This affirmation is true 
because of the rotation invariance of the Wigner function.
Indeed, the Wigner function of a statistical mixture of Fock states is rotation
invariant being a sum of function of the form $\mathcal{L}_n(r^2)$. The 
converse is also true: from the fact that the set of functions $(\mathcal{Y}_{lm}(\theta)
\mathcal{L}_n)$ form a basis of the functions $f(\theta,r)$, a rotation invariant
Wigner function $W(r)$ can be represented as a sum of Laguerre polynomials.
In terms of density operator (Fourier transforming the Wigner function), we end
up with a statistical mixture of Fock states.}.

\begin{figure}
	\centering
	\includegraphics{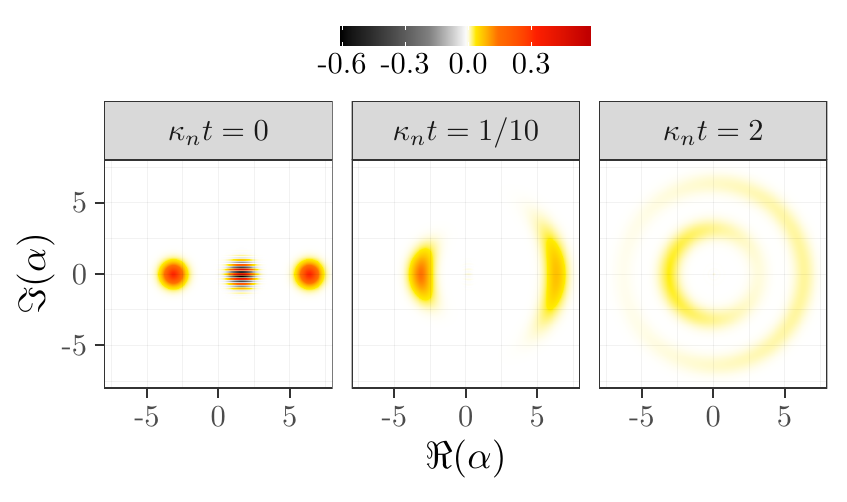}
	\caption{%
		Depending on the preparation, with for instance here
		$\alpha_+ = \sqrt{40}$ and
		$\alpha_- = -\sqrt{10}$, the interference spot is totally 
		washed out by the $L_n$ channel ($\kappa_a = 0$) after
		the spreading timescale 
		$\kappa_n t= 1/10$ leading at longer times
		$\kappa_n t= 2$ to a mixture of Fock states.
	}
	\label{fig:wigner_fockorth}
\end{figure}

In fact, we can better understand the structure of the Wigner
function at finite times by explicitly writing the periodicity in the
variable $\theta$ hidden in \cref{eq:wigner:fockdeco}. Indeed,
we have
$
	W(r,\theta,t)
	=
	\int_{0}^{2\pi}
	W_0(r,\theta') \, \vartheta\left(
		\frac{\theta - \theta'}{2 \pi}
		;
		\mi \frac{\kappa_n t}{2 \pi}
	\right)
	\, \frac{\md \theta'}{2 \pi}
$
with $\vartheta(z;\tau) = \sum_{n \in \mathbb{Z}}
\exp(\pi \mi n^2 \tau + 2 \pi \mi n z)$ the Jacobi theta function.
When $\kappa_n t \gg 1$, the expansion
$
	\vartheta\left(
		\frac{\theta}{2 \pi}; \mi \frac{\kappa_n t}{2 \pi}
	\right)
	\simeq
	1 + 2 \sum_{n \in \mathbb{N}^*}
	\left(\me^{- \kappa_n t/2}\right)^{n^2}
	\cos n \theta
$
can be used to approximate the Wigner function. For instance, if
$\kappa_n t = 2$, only the first harmonic of the Wigner function
dominates with oscillations in amplitude of \SI{74}{\percent} of 
the average value, the second harmonic being less than \SI{2}{\percent}, as
can be seen on \cref{fig:wigner_fock,fig:wigner_fockorth}.
This harmonic decomposition also shows that in the presence of
symmetries in angle, for instance 
$W(r,\theta + 2\pi/p) = W(r, \theta)$ 
for an integer $p \ge 2$, the first non-zero modulation term scales as
$\me^{-\kappa_n t p^2/2}$. Thus, symmetric initial states will 
decohere much quicker than the ones that are not.

Finally, note that the presence of an interference pattern on \cref{fig:wigner_fock} is not
a signature of coherence between Fock state. For a different preparation like the one 
on \cref{fig:wigner_fockorth}, the interference spot is completely washed out over 
a timescale of $\tau_s$.

\section{Einselection process}
\label{sec:einselection}

\begin{figure}
	\centering
	\includegraphics{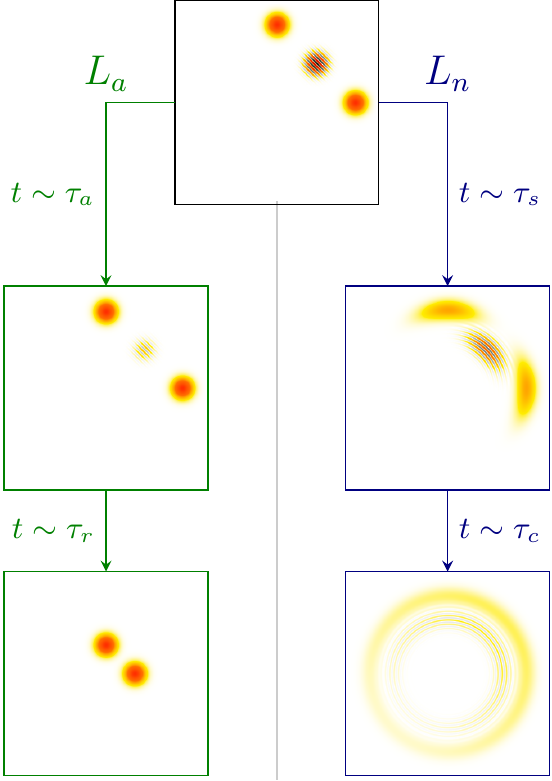}
	
	\caption{%
		Summary of different evolution and timescales of a superposition of coherent
		states under the effects of two decoherence channels: the $L_n$ channel
		induces a spreading of the wavepacket while the $L_a$ channel destroys
		interferences and makes the system relaxes to the vacuum.
	}
\label{fig:summary_effects_timescales}
\end{figure}

Now that we have a proper analytic solution of the problem
and that we completely understand the evolution and characteristic
timescales of each decoherence channel separately, which is 
summarized in \cref{fig:summary_effects_timescales}, we can
analyze the einselection process in its full generality.

\subsection{Pointer states dynamics}

Intuitively, we expect the classical picture that emerges must be 
dominated by the quantum channel which has the strongest coupling constant
\cite{Novais-2005}. We can in fact play with two parameters, the ratio
$\kappa_n / \kappa_a$ and the average energy of the state $\langle n \rangle$.

One regime is when we have 
$\tau_a / \tau_c \approx \kappa_n / \kappa_a \langle n \rangle \gg 1$ or
equally $\langle n \rangle \ll \kappa_n / \kappa_a $. We expect
in this case that Fock state decoherence dominates the dynamics. We can refine
our statement by unraveling two subregimes with respect to the relaxation
timescale $\tau_r$:
\begin{itemize}
\item In the range $\tau_c \ll \tau_a, \tau_r$, or equally 
$\langle n \rangle \ll \kappa_n / \kappa_a$ and 
$\tau_r/\tau_c \approx \kappa_n / \kappa_a \gg 1$,
we have a proper decoherence on the Fock basis as we expected.
\item However, for $\tau_r \ll \tau_c  \ll \tau_a$, or equally 
$\langle n \rangle \ll \kappa_n / \kappa_a$ and 
$\tau_r/\tau_c \approx \kappa_n / \kappa_a \ll 1$, 
we end up in the almost degenerate case $\langle n \rangle \ll 1$ 
where we relax on the vacuum state.
\end{itemize}
Thus, in the regime $\tau_a \gg \tau_c$ where the Fock decoherence
dominates initially, we see that the relaxation induced by the $L_a$ channel 
forbids to transition toward a proper non-trivial coherent-state 
classical picture: before evolving towards a proper statistical mixture of 
coherent states, the system relaxes to the vacuum. All in all, the emergent
physically meaningful classical picture is given by Fock states.

\begin{figure}
	\centering
	\includegraphics{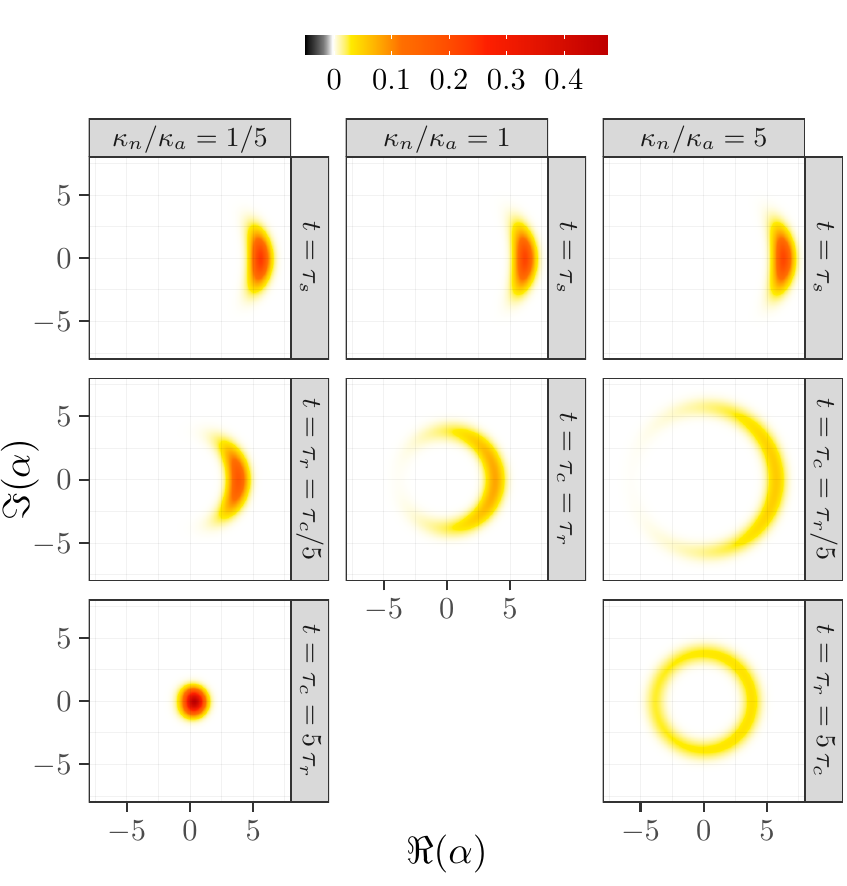}
	\caption{%
		Coherent state dynamics for $\alpha = \sqrt{40} \gg \kappa_n/\kappa_a$. 
		Depending on the relative 	ratio of the coupling constants 
		$\kappa_n/\kappa_a$ we either see a emergent classical picture based solely
		on coherent states $\kappa_n/\kappa_a < 1$ or both coherent and then 
		Fock states ($\kappa_n/\kappa_a > 1$).%
	}
	\label{fig:coherent_dyn}
\end{figure}

From the einselection perspective, a richer dynamics can be found in the
complementary regime where
$\tau_a / \tau_c \approx \kappa_n / \kappa_a \langle n \rangle \ll 1$ or
equally $\langle n \rangle \gg \kappa_n / \kappa_a $. 
\Cref{fig:coherent_dyn} shows the evolution of the Wigner function of 
a coherent state (with initially 40~photons on average) for this situation. 
We can at least see three different regimes: 
\begin{itemize}
\item The regime with $\kappa_a \gg \kappa_n$ (first column with 
$\kappa_a = 5\,\kappa_n$). In this case, we see that coherent states
remain largely unaffected by the environment and evolves
according to the dynamics of the $L_a$ channel with the characteristic
dissipation timescale $\tau_r$.
\item The complementary regime with $\kappa_n \gg \kappa_a$ (third column with 
$\kappa_n = 5\,\kappa_a$) where the $L_n$ rules the dynamical evolution.
We see that the coherent state is spread into a statistical mixture of
Fock states over a timescale $\tau_c$ which is then followed by the much
slower process of relaxation.
\item The intermediate regime with $\kappa_a \approx \kappa_n$. In this case,
it is not possible to clearly conclude which basis is the most classical
one.
\end{itemize}

The previous discussion focused mainly on a preparation of coherent
states. A similar discussion can be made if we prepare a Fock state.
As expected, if the $L_n$ channel dominates, we observe a decoherence
over the Fock basis with a slow relaxation towards the vacuum induce
by the $L_a$ channel. Nonetheless, some subtleties, detailed
in \cref{app:coherent}, occur in the opposite situation because coherent
states do not form a proper orthogonal basis. In the same way that
a very close superposition of coherent states will not properly decohere
under the influence of the $L_a$ channel, a Fock state, being a continuous
superposition of coherent states, cannot properly evolve toward
a classical mixture of coherent states.

\subsection{Approximate pointer states}

For a general dynamics, exact pointer states, defined as states
that do not get entangled with the environment if they are initially
prepared, do not exist. Instead, we have to rely on an approximate
notion of pointer states to give a meaningful notion of an emergent
classical description. Approximate pointer states are defined as
the states that entangle the least with the environment according
to a given entanglement measure. This definition of approximate
pointer states, called the predictability sieve and used in different
contexts~\cite{Kiefer_2007,Riedel-2016-quantum}, is one definition
among different proposal to formalize the idea of robustness 
against the environment. For instance, the Hilbert-Schmidt
robustness criterion~\cite{Gisin_1995,Busse_2009} defines
pointer states to be time-dependent pure states that best 
approximate according to the Hilbert-Schmidt norm the
impure state generated by infinitesimal time evaluation.
Fortunately, those different approaches were shown to be
consistent for physically meaningful models~\cite{Diosi_2000}.

Many different measures of entanglement exist in quantum information
theory and it is not yet totally clear which one is the proper one to
use in the predictability sieve approach. For the sake of simplicity, we
will consider the purity $\gamma$ defined as~\footnote{The purity can 
be used to define a notion of entropy $S = 1-\gamma$ called the linear 
entropy.}:
\begin{equation}
	\gamma = \trace \rho^2
\,.
\end{equation}
Approximate pointer states are then defined by searching
for pure states that minimize the initial variation of the entropy or,
in our case, maximize the variation of purity. It's worth noting that the
purity variation, for a pure state, is also equal to twice the variation of
the principal eigenvalue of the density matrix.

Before proceeding,  we could inquire about the dependence of the approximate
pointer states that we find on the entanglement witness that we choose.
For instance, we could have chosen a whole class of entropies $S_\alpha$
called the Rényi entropies~\cite{renyi1961measures} defined as
$
S_\alpha(\rho) = \frac{1}{1-\alpha} \trace \rho^\alpha
$.
Using those entropies to find approximate pointer state do not
change the conclusion if we are looking at the short-time evolution of
a pure state. Indeed, if we prepare at $t=0$ a pure state, we have that
$\rho^\alpha = \rho$ for $\alpha \ne 0$.
Then $\dot{S}_\alpha = \frac{\alpha}{1-\alpha} \trace \dot{\rho}\rho^{\alpha-1}
=\frac{\alpha}{1-\alpha} \trace \dot{\rho}\rho$ which is equal, apart from
a proportionality factor, to the evolution of purity. The approximate pointer
states do not then depend on which measure we choose.

Coming back to the purity, its derivative $\dot\gamma$,
which is, up to a constant factor, just the derivative of the
largest eingenvalue for an initial pure state, is
linear in $\partial_t \rho$. We thus have $\dot\gamma = \dot\gamma_a +
\dot\gamma_n$, where $\dot\gamma_a$ (resp. $\dot\gamma_n$) is the
contribution solely due to the $L_a$ (resp. $L_n$) channel. 
First of all, since each of these contributions is non positive, a pure state 
will stay pure if and only if it stays pure for each channel. In this case, 
it is easy to see that, unless one of the two constants $\kappa_a$ or 
$\kappa_n$ is zero, the only state staying pure is the vacuum state.
That's why, in general, no exact pointer states exist for a complex
open quantum dynamics and we have to look for approximate ones.
Having said that, we now have to search for pure states that maximize the
evolution of the purity at initial time. As a function of the matrix elements
$\rho_{mn}$ of the initial state written in the Fock basis, the purities 
satisfy the equations:
\begin{subequations}
\begin{align}
	\dot\gamma_n &= - \kappa_n \sum_{m,n} |\rho_{m,n}|^2 (m-n)^2\,, \\
	\dot\gamma_a &= - \kappa_a \sum_{m,n} \Big[
		|\rho_{m,n}|^2 (m+n) \nonumber \\
		&- 2 \rho_{m,n}^* \rho_{m+1,n+1} \sqrt{(m+1)(n+1)}
	\Big]\,.
\end{align}
\label{eq:purity:eqdiff}
\end{subequations}

\begin{figure}
	\centering
	\includegraphics{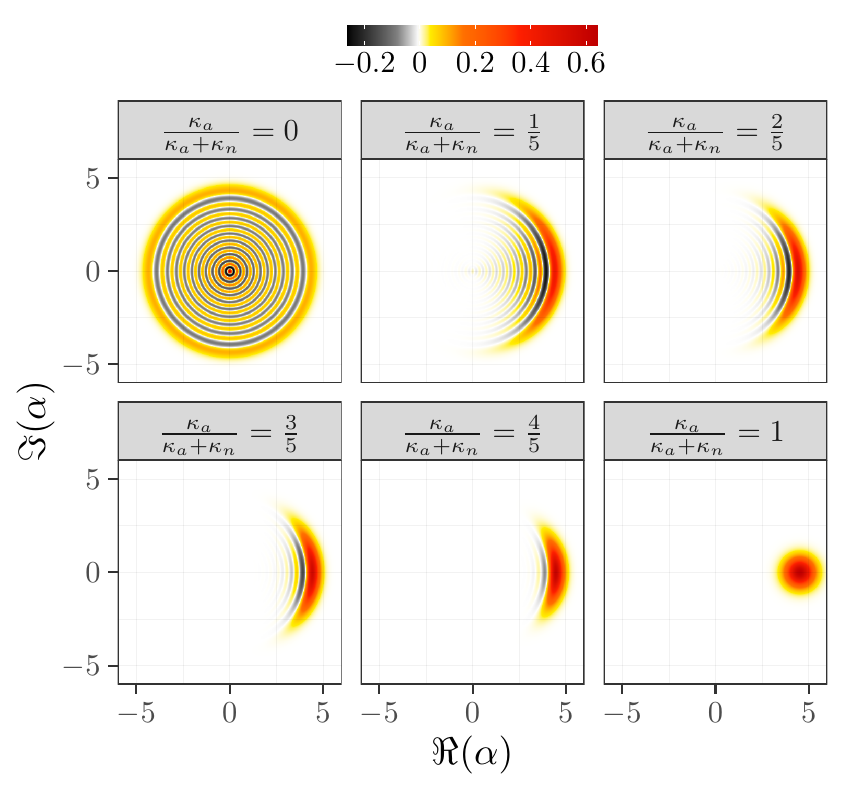}
	\caption{%
		Wavepacket that minimizes the loss of purity at initial time,
		for an average energy of $20$~photons, obtained by
		numerical optimization techniques. The optimal (with respect
		to the purity) state $\dot{\gamma}_{\text{opt}}$ interpolates
		between Fock states and coherent states depending on the
		relative value $\kappa_a/(\kappa_a + \kappa_n)$ of the coupling
		constants.
	}
	\label{fig:optimal_wavepacket}
\end{figure}

\begin{figure}
	\includegraphics{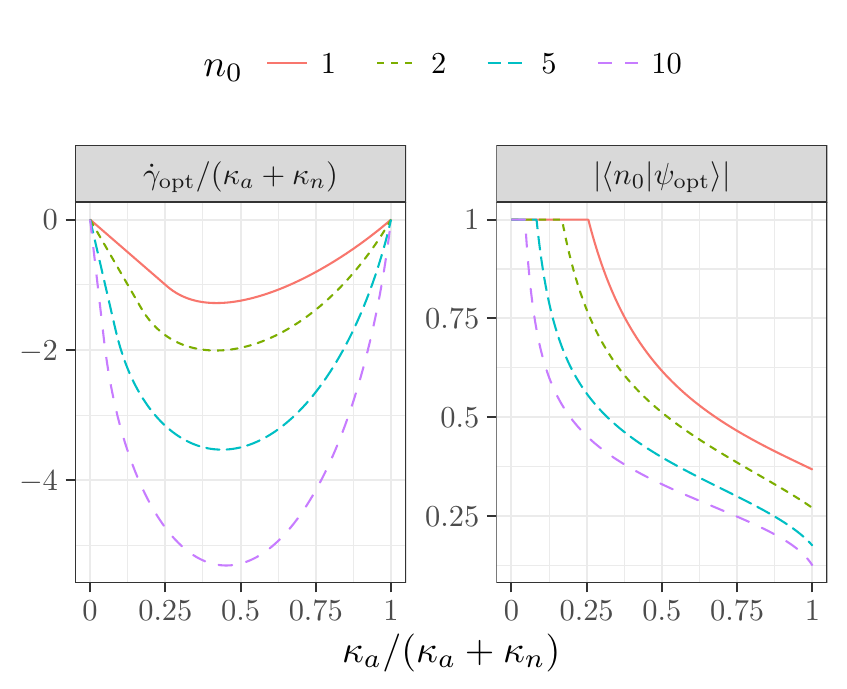}
	\caption{%
		Purity and scalar product of the approximate pointer state
		with Fock states for energies of $n_0 = 1, 2, 5, 10$~photons.
		We see that in the low $\kappa_a$ regime, the approximate
		pointer state $\ket{\psi_\text{opt}}$ is still a Fock state,
		until the critical value $\kappa_n/\kappa_a = n_0 +
		\sqrt{n_0(n_0+1)} + 1/2$. As expected, the variation of purity
		linearly decreases on this plateau.
	}%
	\label{fig:optim:purityandscalprod}
\end{figure}

To find the approximate pointer states, we thus have to optimize the
initial pure state that minimize the absolute purity variation, under
the constraint of a unit norm. It is also natural to fix the average
energy of the wavepacket (otherwise, the vacuum is a trivial optimum).
Since it is difficult to perform this optimization analytically, we
performed it numerically using the Pagmo library~\cite{pagmo2} with
the SNOPT algorithm~\cite{gill2005snopt}. We see on
\cref{fig:optimal_wavepacket} that, when the average energy is a multiple of the
number of photons, the optimal state deforms from a coherent state to a Fock
state. The optimal purity variation,
$\dot{\gamma}_{\text{opt}}$ as a function of $(\kappa_a,\kappa_n)$, can
be seen for different number of photons on
\cref{fig:optim:purityandscalprod}.

\begin{figure*}
	\centering
	\includegraphics[width=0.8\textwidth]{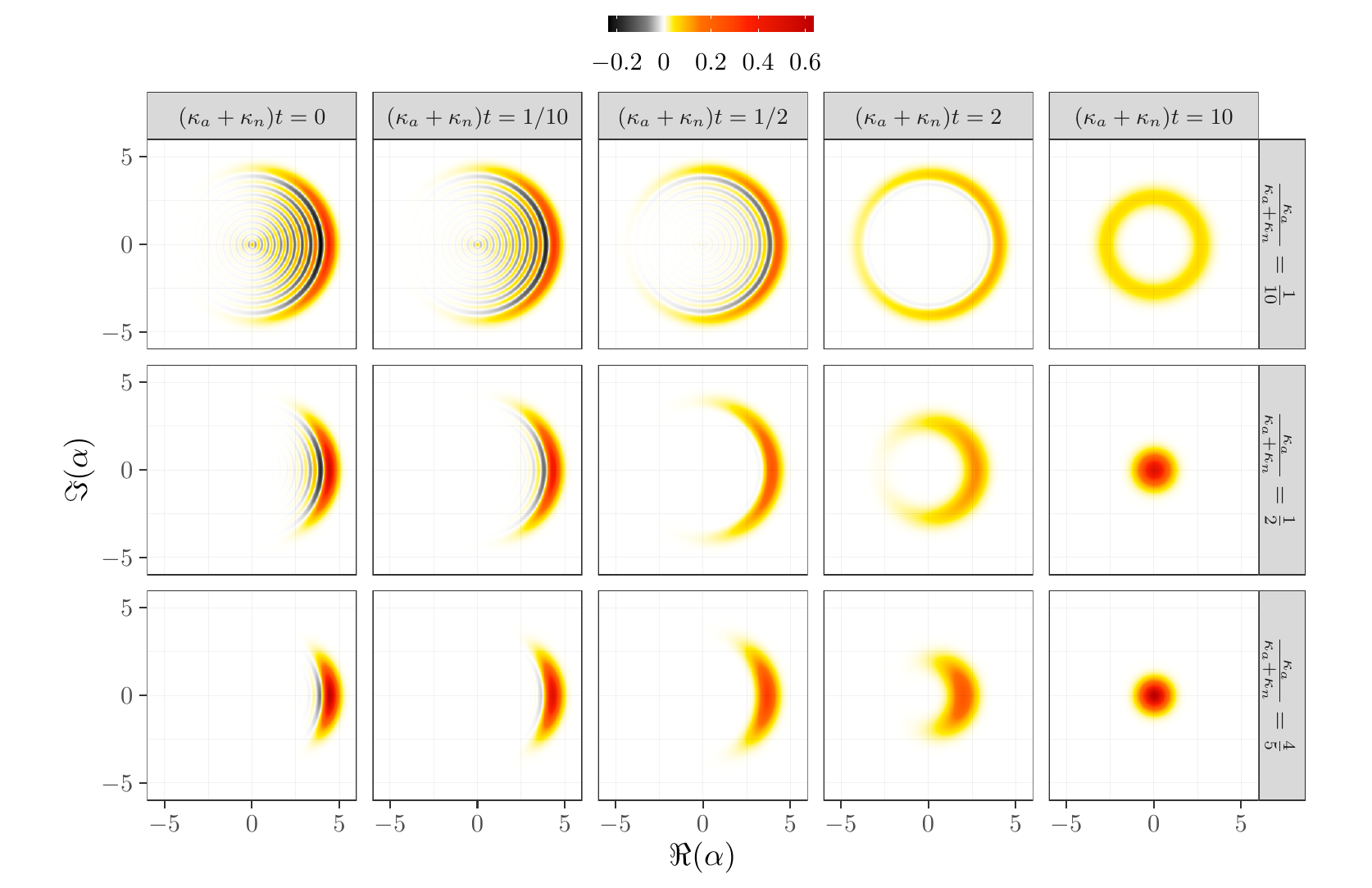}
	\caption{%
		Evolution of the approximate pointer states for several
		$(\kappa_a, \kappa_n)$ parameter regimes, for an initial energy of
		$n_0 = 20$~photons.
	}
	\label{fig:evol_pointers}
\end{figure*}

The time evolution of the approximate pointer states is featured on
\cref{fig:evol_pointers}. This illustrates that on short timescales, the
optimal state keeps its shape and, as such, is quite robust to both
interactions.

Furthermore, a remarkable fact can be seen by looking at the
overlap between the Fock state of energy $n_0$ and the optimal
pure state $\ket{\psi_{\text{opt}}}$: Fock states remain the
approximate pointer states even in the presence of the $L_a$
decoherence channel at small coupling $\kappa_a \ll \kappa_n$.
This is characterized by the presence of a plateau on the overlap
$|\langle n_0 | \psi_{\text{opt}} \rangle|$ as a function of
$\kappa_a/(\kappa_a + \kappa_n)$ as shown on 
\cref{fig:optim:purityandscalprod}.

To better understand this phenomenon, it is instructive to look 
at the evolution of the purity for a state not far away from the Fock 
state $\ket{n_0}$ and see how this small perturbation changes or not
the optimal problem. Assuming that the state $\ket{\psi} = \sum_n c_n \ket{n}$
has coefficients $c_n = \delta_{n,n_0} + \epsilon_n$, with 
$|\epsilon_n| \ll 1$, we have:
\begin{align}
	\dot\gamma_n &= - 2 \, \kappa_n \sum_{n} |\epsilon_n|^2 (n-n_0)^2\,, \\
	\dot\gamma_a &= - 2 \, \kappa_a \Big(
		n_0(1+2\Re \epsilon_{n_0})^2
		+
		\sum_n |\epsilon_n|^2 (n + n_0)
		\nonumber \\
		 &- |\epsilon_{n_0 + 1}|^2 (n_0+1)
		- |\epsilon_{n_0 - 1}|^2 n_0
		\nonumber \\
		&- 2 \Re \epsilon_{n_0+1} \epsilon_{n_0-1}^* \sqrt{n_0(n_0+1)}
	\Big)\,.
\end{align}
The norm and energy constraints can be rewritten as the equation system:
\begin{subequations}
\begin{align}
	2 \Re \epsilon_{n_0} + \sum_n |\epsilon_n|^2 &= 0\,,
	\label{eq:optimization:fock:c1}
	\\
	\sum_n (n - n_0) |\epsilon_n|^2 &= 0
	\,.
	\label{eq:optimization:fock:c2}
\end{align}
\end{subequations}
Using the global phase symmetry of the quantum state, we can fix
$\epsilon_{n_0}$ (an imaginary part on $\epsilon_{n_0}$ corresponds to
changing the phase of the initial Fock state). Furthermore, using both
constraints, we can eliminate all the terms containing $\epsilon_{n_0}$.
The variation of purity takes the form:
\begin{align}
	\frac{\dot\gamma}{2} &=  \kappa_a n_0
	- \sum_n |\epsilon_n|^2 \kappa_n (n-n_0)^2
	\nonumber \\
   &+
   \kappa_a \Big(
		 |\epsilon_{n_0 + 1}|^2 (n_0+1)
		+ |\epsilon_{n_0 - 1}|^2 n_0
		\nonumber \\
		&+ 2 \Re \epsilon_{n_0+1} \epsilon_{n_0-1}^* \sqrt{n_0(n_0+1)}
   \Big)
\,.
\end{align}
We see that under the normalization constraint, Fock states are always
stationary points of the purity variation. Furthermore, provided that $\kappa_n > 4
\kappa_a$, the purity variation is a maximum along all directions except
possibly on the plane $(\epsilon_{n_0-1}, \epsilon_{n_0+1})$. We can
thus concentrate on this plane. In this case, the constraint
\cref{eq:optimization:fock:c2} is simply $|\epsilon_{n_0-1}|^2 =
|\epsilon_{n_0+1}|^2$. If we denote $\theta$ the phase difference
between $\epsilon_{n_0-1}$ and $\epsilon_{n_0+1}$, the tipping point
occurs when:
\begin{equation}
	\kappa_n/\kappa_a  = n_0 + \frac{1}{2}
	+ \sqrt{n_0(n_0+1)} \cos \theta\,.
\end{equation}
Since $\kappa_n$ has to be maximum, we can keep $\theta = 0$. And thus:
\begin{equation}
	\kappa_n/\kappa_a  = n_0 + \frac{1}{2}
	+ \sqrt{n_0(n_0+1)}
	\approx 2 \left(n_0 + \frac{1}{2}\right) \,.
\end{equation}
This equation gives the critical point where Fock states are no
longer the einselected states of our dynamics and explains the 
qualitative features of \cref{fig:optim:purityandscalprod}.
Actually, what we just shown is that a Fock state is always a stationary
point of the purity variation. Given the norm and energy constraints, 
it moves from an extremum to a saddle point exactly when 
$\kappa_n/\kappa_a = n_0 + \sqrt{n_0(n_0+1)} + 1/2$, with 
$n_0$ the average energy of the state. When $n_0 \gg 1$, 
$\kappa_n/\kappa_a \approx 2 n_0 + 1$. As such, when $n_0$ becomes 
bigger, the size of the plateau on which the Fock state remains the
einselected state becomes smaller. Away from the critical point, the
most robust state becomes the state that interpolate between a number
and coherent state of \cref{fig:optimal_wavepacket}.

\begin{figure}
	\includegraphics{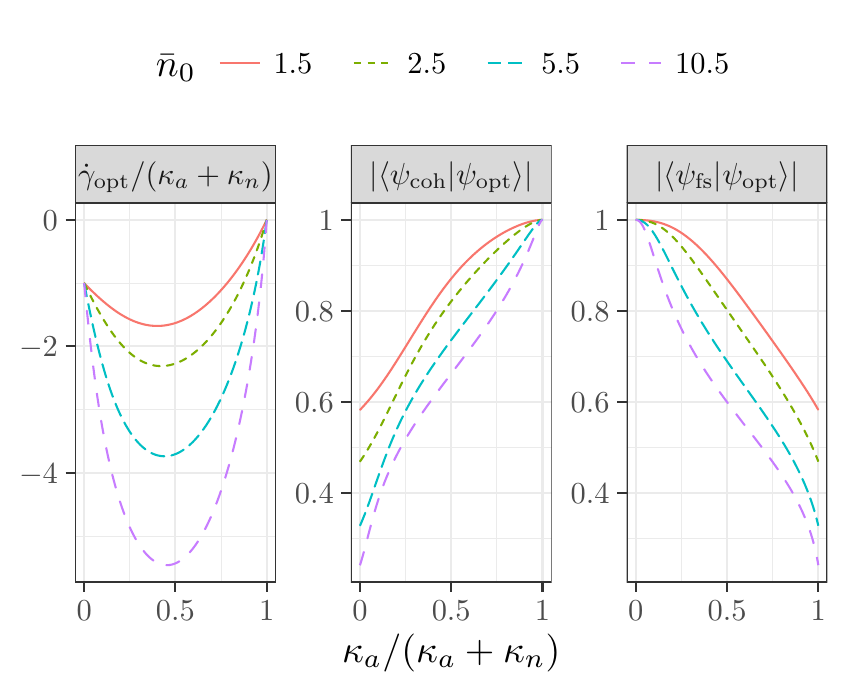}
	\caption{%
		Purity variation, and scalar product of the approximate pointer state
		with the approximate pointer states when one of the coupling
		constant is zero. The average energy corresponds to $\bar{n}_0 =
		1.5, 2.5, 5.5, 10.5$~photons.
		In this case, the optimal state when $\kappa_a = 0$ is
		$\ket{\psi_{\text{sf}}} = (\ket{\lfloor \bar{n}_0 \rfloor} +
		\ket{\lceil \bar{n}_0 \rceil})/\sqrt{2}$. Conversely, when
		$\kappa_n = 0$, we have a coherent state whose average energy is
		$\bar{n}_0$.
		Contrarily to the integer case, there is no plateau appearing in
		the low $\kappa_a$ regime.
	}%
	\label{fig:optim:purityandscalprod:half}
\end{figure}

However, when the average number of photon $\bar{n}_0$ in the cavity is
not an integer, the behavior is quite different. First, even when
$\kappa_a = 0$, there is no exact pointer state satisfying this
constraint. The approximate pointer states in this case are
superpositions of Fock
states $\ket{\psi_{\text{sf}}} = \alpha \ket{\lfloor n_0 \rfloor} +
\beta \ket{\lceil n_0 \rceil}$ with $(\alpha,\beta)$ such that the
energy constraint is satisfied.  Another difference is that in this
case, the optimal state deforms smoothly from the superposition
$\ket{\psi_{\text{sf}}}$ to the coherent state. Contrarily to the
integer case, there is no tipping point, as can be seen on
\cref{fig:optim:purityandscalprod:half}.
\section{Discussions and conclusion}
\label{sec:disc}

\begin{figure}
	\includegraphics{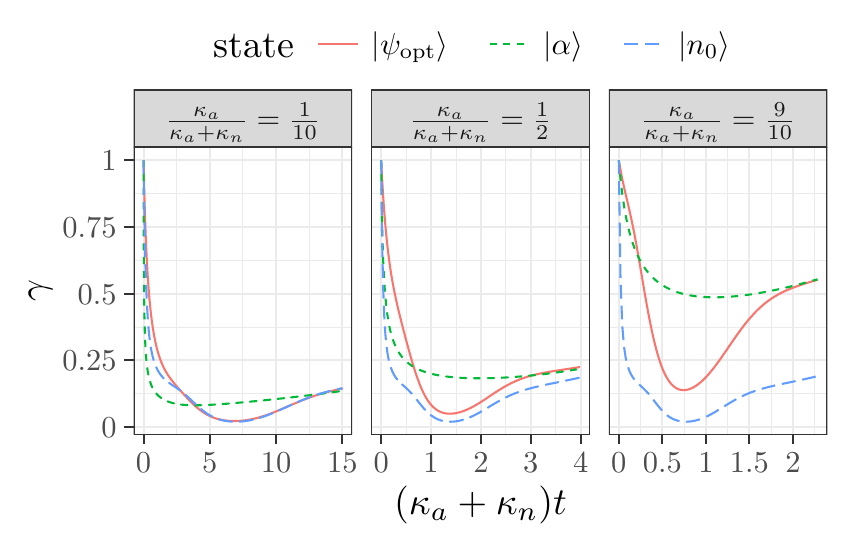}
	\caption{Evolution of the purity for the coherent state, the Fock
		state and the approximate pointer state for the energy $n_0 =
		20$~photons.
	}%
	\label{fig:purity:evolution}
\end{figure}

We discussed here the einselection process in the presence of two incompatible
decoherence channels and used the characterization of approximate pointer
states as states that entangle the least with the environment. By choosing an
entanglement measure, they are found by solving an optimization problem from
the short-time open evolution of pure states (under natural constraints).
Two drawbacks of this approach to the einselection problem can be stated:
how the choice of the entanglement measure influences the answer and does
an optimal state remain robust over time (validity of the short-time hypothesis)?
As we already discussed, focusing on the short-time evolution basically solves 
the issue of which entanglement measure to choose, but the problem remains
open in general.

From the exact solution of the model, we can compare the evolution over time
of the purity between the optimal, the Fock and the coherent state of a given
energy as shown in \cref{fig:purity:evolution}. As expected, we observe
that, at short times, the evolution of purity is the slowest for the
optimal state (by construction)
while, at very long time compared to the relaxation time, everything converges
toward the same value since we are basically in the vacuum. However, at
intermediate timescales, the evolution of purity gets quite involved and we see
that it may even be possible that the optimal state does not remain so and this
is strongly dependent on the coupling constants. In this case,
one can get an intuition of this behavior because the optimal states
have a smaller spreading on the Fock basis than the coherent states and
as such, contract to the vacuum slower. Thus, while the definition of
approximate pointer states is physically intuitive, the question remains
on how to properly characterize them from an information perspective but
also from a dynamical perspective.

A general question that we can also ask in the perspective we adopted here
on understanding the emergence of a classical picture from a complex environment
would be the following: what can we learn about the dynamics and the einselection 
process of the model through the general features of the model
like the algebraic relations between the jump operators? Naturally, the whole problem
depends on the kind of dynamical approximations we do, if we start from the
exact Hamiltonian dynamics or from an approximate master equation like the 
Lindblad equation as we did and still have in mind here. 

Focusing on the general dynamics first, we saw that, given
the Lindblad equation and the commutation relations of the jump operators,
the dynamics can be exactly solved algebraically by adopting a quantum channel
perspective: the full dynamics of the model  decouples and we can look
at the evolution of both quantum channels separately, which is an easy
task. In fact, our solution can be abstracted in the following sense.
Consider an open quantum dynamics with a damping channel encoded
by the jump operator $L$ and a number decoherence channel encoded 
by the jump operator $L^\dagger L$. Our approach can then be followed steps
by steps again. It however opens the question on how to defined a proper
phase space associated to a jump operator (through, for instance, generalized
coherent states \cite{Perelomov-2012}), how to generalize characteristic functions
like the Wigner representation and verify if the phase space perspective on decoherence
developed throughout our analysis still holds in this generalized context.

Concerning the einselection process however, note that having a general solution
does not mean that finding the approximate pointer states is also solved. 
The problem remains here largely open when we have many
incompatible decoherence channels. Indeed, not only does the algebraic
structure influences the einselection process but also the set of coupling
constants and how they run as a function of the energy~\cite{Neto-2003,Novais-2005}.

We can start to grasp those subtleties already when a thermal environment is 
present. Indeed, absorption processes can occur
which forces us 
to consider a jump operator of the form 
$L_{a^\dagger} = \sqrt{\kappa_a\bar{n}} \, a^\dagger$. However,
such a jump taken alone is hardly meaningful (the state evolves toward an infinite
energy configuration). Relaxation processes have to be taken into account and
this is summed up by the usual commutation relations between $a$ and $a^\dagger$.
Still, this is not sufficient to reach equilibrium and proper relations between the 
coupling constants of the different processes must exist. This well known example
already shows that relating the general features of the open quantum dynamics 
to the einselection process is not that straightforward.

The model studied in this paper behaves quite intuitively: when one of the 
coupling dominates, the associated decoherence channel controls the 
einselection process. However, when both couplings are comparable, 
the non-trivial commutation relations between the jump operators enter the game. 
No exact pointer basis exists and the robust states interpolate between the 
extreme cases and how far they are from them depends on the relative values 
of the coupling constants again.
Still, we also unraveled the fact that, given some constraints, the transition from
one class of pointer states to another as a function of the relative values of the
couplings is not smooth: below a critical value, Fock states remains exactly the
most robust states. How can such a behavior be anticipated from the structure
of the dynamics remains to be explored.
In the end, this shows that the question of predicting general features of the emergent classical
picture only from the structure of the interaction (algebraic relations between jump
operators, set of coupling constants) still calls for a deeper understanding.

In summary, we solved exactly a model of decoherence for an open quantum system
composed of two incompatible decoherence channels using quantum trajectories and
phase-space techniques. We then studied numerically the dynamical
emergence of a classical picture. We were then able to see how the selection of
approximate pointer states depends on the relative values of the coupling
constants. This unraveled the remarkable robustness of Fock states relative to
a decoherence on the coherent state basis and we were able to analyze
quantitatively the critical coupling where this robustness gets lost. Our results
show that the physics of decoherence and einselection still has a lot to offer
when a complex dynamics is at play.

\begin{acknowledgments}
	The authors thank Ekin Ozturk for his help with Pagmo. We also thank
	Pascal Degiovanni for fruitful discussions and his comments on the
	manuscript.
\end{acknowledgments}

\appendix
\section{Details on the Wigner representation}

\subsection{Decoherence on the Fock basis}
\label{app:Fock}

If decoherence occurs on the Fock basis, the components are
modulated by an exponential factor with a decohrence timescale inversely
proportional to the ``distance'' between two Fock states.
 We can then write the Wigner function as:
\begin{align}
	W_t(r,\theta)
	&=
	\sum_{l \in \mathbb{Z}}
	W_l(r) \me^{-\kappa_n l^2 t/2} \me^{\mi l \theta}\\
	&=
	\int_{0}^{2\pi}
	W_0(r,\theta') \, \vartheta\left(
		\frac{\theta - \theta'}{2 \pi}
		;
		\mi \frac{\kappa_n t}{2 \pi}
	\right)
	\, \frac{\md \theta'}{2 \pi}\\
	&=
	\int_{-\infty}^{+\infty}
		W_0(r,\theta') \, \mathcal{N}_{\sigma^2 = \kappa_n t}\left(
		\theta - \theta' 
	\right)
	\, \md \theta'
	\,,
\end{align}
where $W_l(r)$ is the $l$th harmonics of the angle variable of the
Wigner function at time $t=0$, $\vartheta(z;\tau) = \sum_{n \in \mathbb{Z}}
\exp(\pi \mi n^2 \tau + 2 \pi \mi n z)$ is the Jacobi theta function and
$\mathcal{N}_{\sigma^2}$ is the centered normal distribution of variance
$\sigma^2$.

In phase space, the dynamics induced by the $L_n$ channel is thus
the expected Gaussian spreading of a diffusive evolution. When 
$\kappa_n t \ll 1$, the periodicity can be forgotten. However, 
when $\kappa_n t \gg 1$, we have the interesting expansion
of the Jacobi function:
\begin{equation}
	\vartheta\left(
		\frac{\theta}{2 \pi}; \mi \frac{\kappa_n t}{2 \pi}
	\right)
	=
	1 + 2 \sum_{n \in \mathbb{N}^*}
	\left(\me^{- \kappa_n t/2}\right)^{n^2}
	\cos n \theta
\,.
\end{equation}
Consequently, for $\kappa_n t = 2$, only the first harmonics dominates with
amplitude oscillations  of \SI{74}{\percent} of the average value,
as can be seen on \cref{fig:wigner_fockorth}.

Note that, if there is a discrete symmetry in the angle variable of
the Wigner function, such as $W(r,\theta + 2\pi/p) = W(r, \theta)$ for
an integer $p \ge 2$, the first non-zero modulation term scales as
$\me^{-\kappa_n t p^2/2}$. Here, the scaling is not
linear but quadratic. This could be in fact directly recovered
by noting that , the symmetry $W(r,\theta + 2 \pi/p) = W(r,\theta)$ is
equivalent to having $\rho_{k,k+l} = 0$ if $l$ is not a multiple of $p$.
In the end, initial states possessing a symmetry will thus
form a crown much quicker than the ones that don't.

\subsection{Decoherence on the coherent states basis}
\label{app:coherent}

If we initially prepare the cavity in the Fock state $\ket{m}$, then the
$L_n$ dynamics is trivial. As we have seen on \cref{eq:fock:proba}, the
state inside the cavity can be described as a classical mixture of Fock
state, the probability to have the state $\ket{k}$ being given by:
\begin{align}
	p_{k}
	=
	{m\choose k} (1- \me^{-\kappa_a t})^{(m-k)} \me^{-\kappa_a k t}
	\,,
\end{align}
which is a binomial distribution of parameters $m$ and $p =
\me^{-\kappa_a t}$. The mean of such a distribution is $mp$ and its
variance is $\sigma^2 = m p (1-p)$.

\begin{figure}
	\centering
	\includegraphics{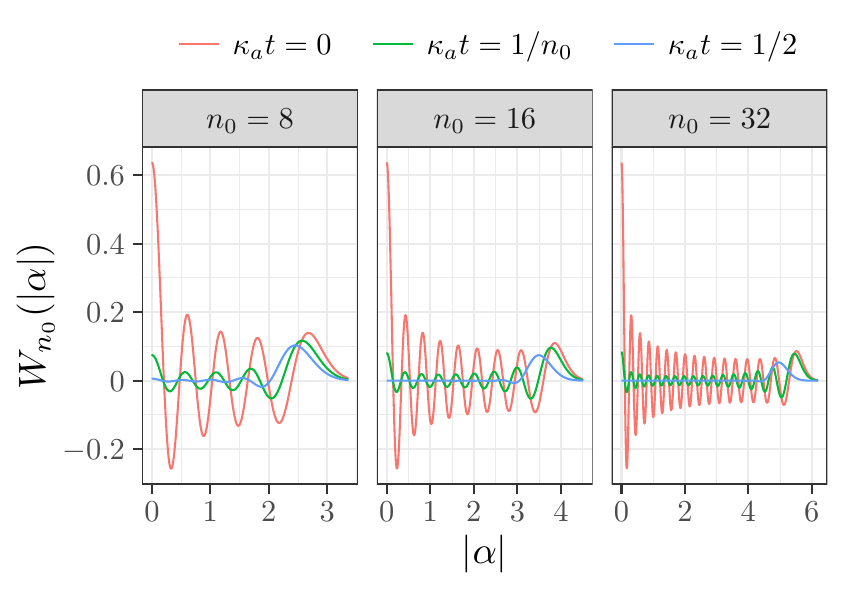}
	\caption{Radial part of the Wigner function for different initial
	Fock states $\ket{n_0}$.}
	\label{fig:fock:radial}
\end{figure}

For small times, we have $\sigma^2 \simeq m \kappa_a t$. This timescale
is comparable to the decoherence timescale $\tau_a$ we introduced in
\cref{eq:timescales:taua} for the coherent states. We see that in this
case, it corresponds to the time it takes for the probability
distribution to spread over several Fock states. In the Wigner function,
it corresponds to the attenuation of the oscillations near the origin
(see \cref{fig:fock:radial}).

On the contrary, this does not correspond to a fast decoherence over
coherent states. Since the Fock states have no definite phase, it is
natural to look for a mixture of coherent states which is uniform in its
phase distribution. A natural expression for any state without phase
preference would thus be:
\begin{equation}
	\rho = \int q(n_0) \ket{\sqrt{n_0} \me^{\mi\theta}}
	\!
	\bra{\sqrt{n_0} \me^{\mi\theta}}
	\, \md \theta \md n_0\,.
\end{equation}
We can easily show that
\begin{equation}
	\rho = \sum_n
	\int q(n_0) P_{n_0}(n) \, \md n_0
	\ket{n}\!\bra{n}\,,
\end{equation}
where $P_{n_0}$ is the Poisson distribution of rate $n_0$. The resulting
distribution over Fock states thus has a variance equal to its mean.
This is approximately the case for the binomial distribution only when
$p \simeq 0$.

As such, the timescale to have decoherence over coherent states from an
initial Fock states is $1/\kappa_a$, the same as the relaxation scale.
This situation is actually similar as the one for coherent states in
respect to the $L_n$ dynamics: the short timescale governs the spreading
in phase space, while the long timescale governs the decoherence between
the Fock states. This is summarized on \cref{table:timescales}.

\bibliography{biblio}

\end{document}